# Optimization of Battery Energy Storage to Improve Power System Oscillation Damping

Yongli Zhu, *Student Member, IEEE*, Chengxi Liu, *Member, IEEE*, Kai Sun, *Senior Member, IEEE*, Di Shi, *Senior Member, IEEE*, Zhiwei Wang, *Senior Member, IEEE*

*Abstract*—This paper studies the optimization of both the placement and controller parameters for Battery Energy Storage Systems (BESSs) to improve power system oscillation damping. For each BESS, dynamic power output characteristics of the power converter interface are modelled considering the power limit, State of Charge limit, and time constant. Then, a black-box mixed-integer optimization problem is formulated and tackled by interfacing time-domain simulation with a mixed-integer Particle Swarm Optimization algorithm. The proposed optimization approach is demonstrated on the New England 39-bus system and a Nordic test system. The optimal results are also verified by time-domain simulation. To improve the applicability and efficiency of the proposed method, seasonal load changes and the minimum number of BESS units to be placed are considered. The proposed controller is also compared to other controllers to validate its performance.

## I. INTRODUCTION

POWER system oscillation at a low frequency in the range of 0.2 to 2.5 Hz typically happens in interconnected power systems with weak tie-lines [1]. Traditionally, oscillation can be mitigated by fine-tuning the Power System Stabilizer (PSS) with each involved generator. However, for large interconnected power systems, such control requires a coordinated parameters-tuning scheme for many generators. This will inevitably involve different regulation entities and necessitate real-time reliable information shared among different control regions. Moreover, many researchers suggest using a centralized control system for online PSS tuning. However, this will also bring the challenges of time delays and communication costs among multiple interconnected regions. One alternative way is to use local FACTS devices such as SVC, TCSC, and STATCOM to offer extra damping support [2]-[4]. Basically, these devices are either passive elements or sources that alleviate system oscillations by controlling reactive power or varying the line admittances. Using renewable energy like wind power to provide extra inertia for oscillation damping or frequency regulation has also been considered [5]-[7].

With the rapid development of battery technology and power electronic converters, more utility-scale Battery Energy Storage Systems (BESSs) have been deployed in power grids and begun to play important roles in grid operations. As an active source, a BESS can be used for load following or balancing as an ancillary service participating in grid operations and power markets [8]-[11]. Under normal conditions of a power system, a BESS is operated at its steady state, i.e. either charging or discharging in a scheduled mode. However, due to considerable initial investment of a BESS, its function can be further exploited to, e.g., helping to improve the damping against system-wide oscillations.

Many research activities about energy storage control to improve power system stability have been reported. Papers [12] and [13] propose a control method to increase the damping ratio of a target mode to a desired level by energy storage. In [14] and [15], robust damping controllers are designed for multiple Superconducting Magnetic Energy Storage devices in a multi-machine system by solving a constrained Min-Max optimization problem or a Linear Matrix Inequality (LMI) optimization problem. Paper [16] proposes a Particle Swarm Optimization (PSO) based oscillation damping controller optimized by heuristic dynamic programming and tested it on a two-area system with one energy storage device. Paper [17] proposes a damping controller based on a STATCOM equipped with energy storage. Paper [18] designs a damping controller based on proposed damping-torque indices. Ref. [19] proposes an anti-windup compensator for energy storage-based damping controller. Paper [20] applies the Port-Hamiltonian method to nonlinear BESS models to improve transient stability.

Besides the controller design for a single BESS device, another challenging problem is the placement of multiple BESSs. Most researchers investigated that problem from the viewpoint of energy management to minimize, e.g., operating costs. For instance, in [21], the optimal siting and sizing problems of multiple BESSs for daily energy management of the distribution network have been studied. In [22], research was conducted for the optimal scheduling and sizing of BESSs in a microgrid by the Vanadium Redox Battery systems.

Most existing studies on energy storage placement have been in the economic or steady-state aspects or at the distribution system level. Few studies have investigated the placement problem from the stability enhancement perspective

This work was supported by the This work was supported in part by NSF CAREER Award under Grant ECCS-1553863, in part by the ERC Program of the NSF and DOE under Grant EEC-1041877 and in part by the SGCC Science and Technology Program under project Hybrid Energy Storage Management Platform for Integrated Energy System.

Y. Zhu, C. Liu and K. Sun are with the Electrical Engineering Department, the University of Tennessee, Knoxville, TN 37996 USA. (e-mail: yzhu16@vols.utk.edu, cliu48@utk.edu, kaisun@utk.edu).

D. Shi and Z. Wang are with GEIRI North America, San Jose, CA 95134, USA. (e-mail: di.shi@geirina.net, zhiwei.wang@geirina.net).

for transmission systems. In [23], the Genetic Algorithm is tested on the IEEE 14-bus system to determine the best sites to install energy storage devices for system voltage stability, whose controller parameters are predefined and not optimized together with the locations. In [24], the controller parameters are optimized by Tabu-Search with the locations fixed. Therefore, the optimal BESS placement problem to improve system oscillation damping has not been studied well.

Compared to existing works, the contributions of this paper include: 1) establishing a simulation-based optimization framework for solving the BESS placement problem, which is more convincing than conventional small signal analysis; 2) co-optimizing the locations and controller parameters for multiple BESS units by a Mixed-integer Particle Swarm Optimization (for short, Mixed-PSO) algorithm; 3) cost analysis on BESS units in the proposed optimization framework. Typically, oscillations regarding inter-area modes are more concerned than local modes in grid operations, and if not damped well, those modes can be extremely harmful to power system stability. The goal of the proposed approach is to help damp a target inter-area mode with a desired damping ratio improvement without worsening the other modes.

In the rest of the paper, section II presents a BESS power output model for oscillation studies. Section III formulates the optimization problem with its objective and constraints. Section IV elaborates the detailed procedure of the proposed simulation-based optimization approach using Mixed-PSO. Then, case studies on the New England 39-bus system and Nordic test system are presented in Sections V and VI. Section VII studies the applicability of the proposed approach and its improvement. Section VIII concludes the paper.

## II. BESS Model for Oscillation Damping Study

As shown in Fig. 1, a BESS typically consists of the storage part, i.e. battery cells, and a Power Conditioning System (PCS), which is typically composed of a DC/DC converter mainly for battery charging/discharging and a DC/AC converter to maintain the pre-specified voltage and power outputs for integration with the AC power grid. A battery cell can be represented by an equivalent voltage source nonlinearly depending on its SOC (State-Of-Charge), which is defined by (1) as the remaining energy divided by its total energy capacity $E_{total}$.

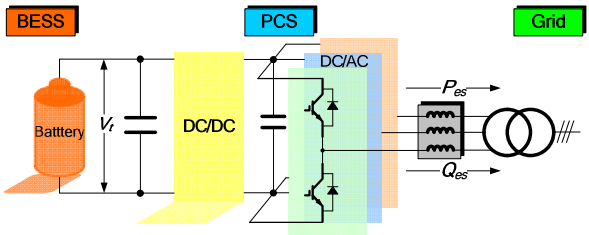

Fig. 1. A typical topology of utility-scale BESS.

$$\text{SOC} = 1 - \Delta E / E_{total}, \quad \Delta E = \int_{t_0}^{t_1} P_{es}(t)dt \quad (1)$$

Most grid-related control strategies are implemented in the DC/AC part, and the cell voltage is maintained at a specific level by the DC/DC part or the battery mange system (BMS). Therefore, in this paper, the BESS will be modeled by its PCS focusing on the DC/AC part but simultaneously considering the nonlinearity of the cell via its SOC. In practice, the battery cell will be protected from deep charging or discharging for life-span considerations. The allowable SOC range is set between $SOC_{min}$ and $SOC_{max}$ in this paper. The equivalent circuit model considering SOC is illustrated in Fig. 2.

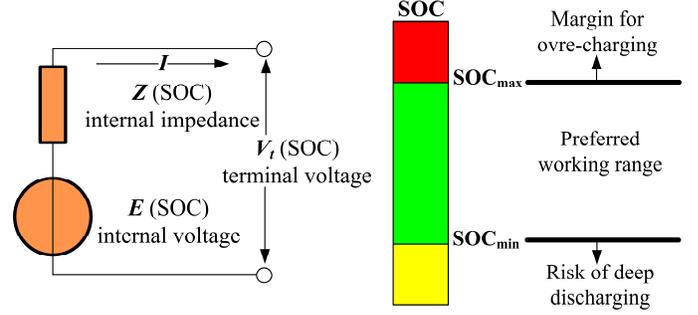

Fig. 2. The equivalent circuit of one battery cell.

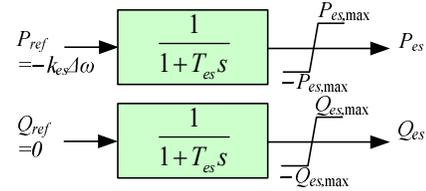

Fig. 3. P-Q decoupled control scheme.

For system oscillation studies, the PCS of a BESS can be approximated by a first order transfer function shown in Fig. 3 like those in [12] and [24]-[26], because in a typical P-Q decoupled control scheme for the PCS, active and reactive powers can be independently regulated to their reference values. Meanwhile, responses of power electronic devices (e.g. switching on or off) are typically much faster than electromechanical dynamics of synchronous generators. The effectiveness of this model for oscillation damping studies has been validated by experiments and analyses in [25] and [26].

$$P_{es} = \begin{cases} \dfrac{P_{ref}}{1+T_{es}s}, & \text{SOC} < \text{SOC}_{max} \text{ and } P_{ref} < 0 \\ & \text{SOC} > \text{SOC}_{min} \text{ and } P_{ref} > 0 \\ 0, & \text{otherwise} \end{cases} \quad (2)$$

$$P_{es} \in [-P_{es,\max}, P_{es,\max}], \quad P_{ref} = -k_{es}\Delta\omega \quad (3)$$

The power output model of the BESS is given by (2) and (3). Because the active power and frequency are more correlated in AC power systems, the proposed damping controller mainly adjusts the active power output (positive when battery discharges) using a frequency deviation signal while maintaining the reactive power output to zero, as shown in Fig. 3. The terminal bus frequency $\Delta\omega$ is used as the input signal for $P_{ref}$. In Fig. 3 and (3), $k_{es}$ (>0) is the controller gain for each BESS and is directly associated with the objective function for optimization in the next section. $T_{es}$ is the time constant of the BESS power converter and is set to 0.02 sec in simulations. Generally, it is much smaller than the inertia time constants of large generators [26]. In addition, the total energy capacity $E_{total}$ of a utility-scale BESS can range from about 2 to 300 MWh [27]. In this paper, a standard capacity of 10MWh is assumed for each BESS.

III. PROBLEM FORMULATION

The overall problem is formulated as follows:

$$\min_{z_i, k_{esi}} Obj = \sum_{i=1}^{N} z_i k_{esi} \quad (4)$$

$$\text{s.t.} \quad \sum_{i=1}^{N} z_i = N_{es} \quad z_i \in \{0, 1\} \quad (5)$$

$$\xi_k \geq \xi_k^* \quad (6)$$

$$\Delta \xi_i = \xi_i - \xi_{i0} \geq 0 \quad i = 1, 2, ..., n \quad (7)$$

$$k_{es,\min} \leq k_{esi} \leq k_{es,\max} \quad i = 1, 2, ..., N_{es} \quad (8)$$

For an acceptable frequency deviation during power system oscillation, $k_{es}$ indicates the theoretically maximum power of the DC/AC converter used for damping improvement, so the objective (4) is to minimize the sum of $k_{es}$ values of all the BESSs, where, $z_i$ is a binary variable equal to 1 if and only if a BESS is placed at the $i$th bus. Respectively, constraints (5)-(8) require that the number of BESS units be equal to the given $N_{es}$, damping ratio $\xi_k$ for a target mode not be less than a threshold (typically 5% in practice), damping ratios of the other electromechanical modes not decrease, and BESS controller gains should be within a defined range.

The values of $k_{es,\max}$ and $k_{es,\min}$ are related to the converter power limits. A larger $k_{es,\max}$ may lead to more conservative optimization results (i.e. larger objective value) and make the algorithm run longer due to the bigger searching region. In this paper, the possible range of the $k_{es}$ value is determined by the following empirical approach: the power rating for most existing utility-scale BESS projects range from about 5 to 100MW [27], i.e. $P_{es,\max} \in [0.05, 1.0]$ p.u. (refer to 100MVA base); also assume $|\Delta\omega| = 0.01$ p.u. Finally by (3), it leads to: $k_{es}=|-P_{es}|/|\Delta\omega| \leq P_{es,\max}/|\Delta\omega| \in [0.05/0.01, 1.0/0.01] = [5, 100]$. The upper limit $k_{es,\max}$ can be further manually adjusted by trial-and-error method depending on the studied systems scale.

IV. SIMULATION-BASED OPTIMIZATION SCHEME

A. Mixed-PSO

The problem defined by (4)-(8) is a non-convex, nonlinear mixed-integer optimization problem. Conventional gradient-based programming algorithms do not apply. A mixed integer-encoding PSO (for short, Mixed-PSO) is introduced here to solve the problem.

*1) Encoding Scheme for Mixed-variables*

In this paper, each decision vector $X_i \in \mathbf{Z}^N \times \mathbf{R}^N$ is composed of variables on locations and control gains, i.e. $X_i = [\mathbf{locs}, \mathbf{k_{es}}]$. For integer variables, there are two schemes for encoding: one is binary-encoding, such as $\mathbf{locs} = [01000...111] \in \{\mathbf{0, 1}\}^N$ with each bit representing the decision of placing a BESS or not; the other is direct-integer-encoding, i.e. $\mathbf{locs} = [loc_1, loc_2, ..., loc_{Nes}] \in \mathbf{Z}^{Nes}$ and $loc_i \in \{1, 2, ..., N\}$. In a large power system with many buses, the number $N_{es}$ of utility-scale BESSs to be placed is typically small. Thus, the searching space by the integer-encoding scheme can be smaller than that of the binary-encoding scheme, e.g. $O(N^{Nes}) < O(2^N)$ for $N \geq 25$ and $1 \leq N_{es} \leq 5$. Therefore, the direct-integer-encoding scheme is adopted here for $\mathbf{locs}$. For $\mathbf{k_{es}}$, the real-number-encoding is used due to its continuous nature.

*2) Updating Formula*

In the PSO, each solution vector is named as a "particle" representing its current "position" in the searching space. To update the solution vector, the so-called "velocity" vector is internally generated by the algorithm to update the "particle position" in each "generation" (i.e. iteration). The "velocity" uses a weighted sum of 1) the previous velocity, 2) the difference between the current and "individual best" positions, and 3) the difference between the current and "group best" positions. For an $n$-particle swarm in a $D$-dimensional space, they are shown in the following equations:

$$V_{id}^{k+1} = \omega V_{id}^k + c_1 r_1 (P_{id}^k - X_{id}^k) + c_2 r_2 (P_{gd}^k - X_{id}^k) \quad (9)$$

$$X_{id}^{k+1} = X_{id}^k + V_{id}^{k+1} \quad i = 1, 2..., n; \; d = 1, 2, ..., D \quad (10)$$

where, $\omega$ is the "inertia" weight, i.e. the portion of the velocity component in the previous generation; $V_{id}^{k+1}$ is the $d$th component of the $i$th speed vector in the $(k+1)$th generation; $X_{id}^{k+1}$ is the $d$th component of the $i$th particle in the $(k+1)$th generation; $X_{id}^k$ is the $d$th component of the $i$th particle in the $k$th generation; $P_{id}^k$ is the $d$th component of the "individual best" position of the $i$th particle in the $k$th generation; $P_{gd}^k$ is the $d$th component of the "group best" particle position found until in the $k$th generation; $r_1$, $r_2$ are random constants drawn uniformly from [0,1]; $c_1$, $c_2$ are fixed non-negative numbers ("accelerating factors") for convergence purpose.

*3) Checking Validity of the Solution (Particle)*

The particle position and speed generated by the PSO should be within a valid range and satisfy certain explicit engineering conditions. For example, the elements of the location vector should be distinct during the optimization after updating. The algorithm given in Table I iteratively replaces duplicated elements by checking the nearest elements for any duplicate and replace it by the nearest different integer.

TABLE I. PSEUDOCODE FOR DUPLICATES REMOVING ALGORITHM

| Algorithm 1: Duplicates replacing for the location vector |
|---|
| 1  **Input**: location vector X = [X$_1$, X$_2$, …, X$_{Nes}$] |
| 2  create a set S := {1, 2, …, N} |
| 3  [*val*, *idx*] := **unique**(X) |
| 4  S := S \ {*val*} |
| 5  create a set O := {1, 2, …, N$_{es}$} |
| 6  O := O \ O[*idx*] |
| 7  **for** *i* := 1 to length(O) |
| 8      k = **argmin**$_k$ |S[k] – X[O[*i*]]| |
| 9      X[O[*i*]] := S[k] |
| 10     S := S \ S[k] |
| 11 **end** |
| 12 **Output**: the updated location vector X without duplicates |

B. Simulation-based Optimization

The proposed simulation-based optimization approach can be implemented by interfacing MATLAB with DIgSILENT: MATLAB is the environment to implement the Mixed-PSO algorithm and DIgSILENT is used as the simulation engine to check constraints and evaluate the objective function. Compared with the linearized methods (e.g. model-based small signal analysis), a merit of this simulation-based optimization is that the damping ratio of any oscillation mode under a disturbance can be calculated directly and accurately from time-domain simulation, which allows sufficient considerations of nonlinearities in power system models, such as the limits with excitation systems and BESS power converters. Via a flexible communication interface, DIgSILENT is called by MATLAB to execute simulation whenever the objective function or constraints need to be checked. Its communication with MATLAB is coded in a DPL (DIgSILENT Programming Language) file containing all necessary data structures fed by MALTAB. A flowchart of the proposed optimization scheme is shown in Fig. 4. Note that two internal interface variables: **ES_Locs** and **ES_k$_{es}$** are pre-defined in the DPL file to modify the grid model dynamically.

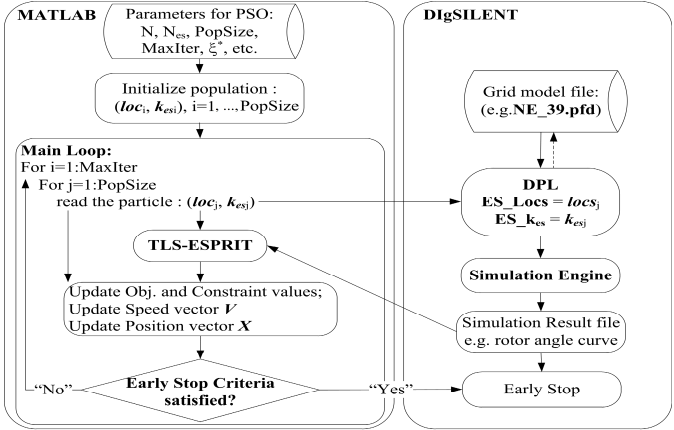

Fig. 4. Flowchart of the simulation-based mixed-PSO optimization scheme.

Regarding the calculation of the modal damping ratio from the simulation curves, the TLS-ESPRIT (Total Lease Square-Estimation of Signal Parameters through Rotational Invariant Technique) method is implemented in MATLAB. Its basic procedure applies SVD (Singular Value Decomposition) twice: the first SVD is on the measurement matrix to extract the modal subspace $U_s$; then the second SVD decomposes $U_s$ to get the optimal estimation for the so-called rotation matrix $\Psi$, whose eigenvalues contain all the modal information of the original signal. The TLS-ESPRIT is reported to be superior to conventional methods like Prony analysis in terms of less sensitivity to noises [28] [29].

## V. CASE STUDY I: NEW ENGLAND 39-BUS SYSTEM

### A. System model and optimization result

The optimal BESS placement algorithm is firstly tested on the New England (NE) 39-bus system shown in Fig. 5. Let the total number $N_{es}$ of BESS units be 3. In DIgSILENT's default model, all inter-area oscillation modes have damping ratios $\xi_k$>5%. A modified model is adopted to reduce the damping ratio of the inter-area mode of 0.657Hz to 1.68% by increasing gains with the AVRs and governors of seven generators. Thus, the goal is to improve the damping ratio of that target mode to at least 5% with the minimum sum of elements in **k$_{es}$**, c.f. (4). The mode rises mainly due to the oscillation between generator groups {g1, g10, g8} and {g9, g5, g7, g6, g4}.

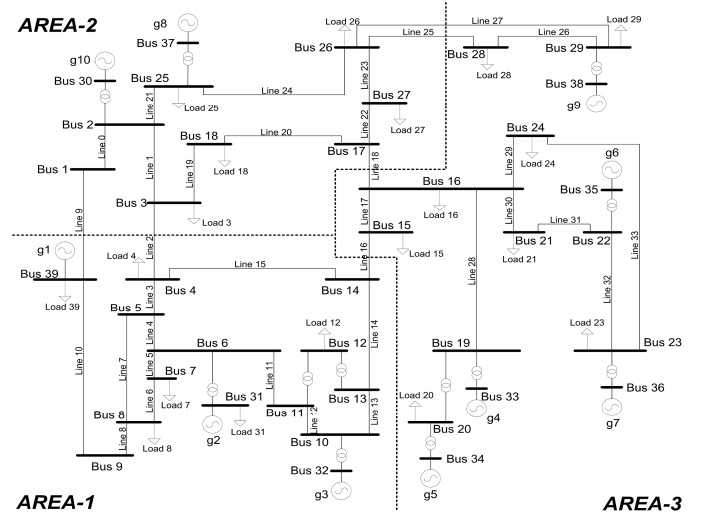

Fig. 5. One-line diagram of the NE 39-bus system.

A three-phase temporary fault added at $t$=0 s and cleared after 0.1s on Bus 16 is simulated for $T$=20 seconds. The fault bus lies on the main system oscillation interface. Parameters of the PSO are: $c_1$=2, $c_2$=2, $\omega$=0.9, $P$=30 (population size) and $I$=30 (maximum iterations), and $k_{es}$=[5, 50].

The optimal location vector and controller gain vector are **locs** = [35, 36, 38] and **k$_{es}$** = [29.5894, 8.2391, 22.5577] with the damping ratio of the target mode at 5.005% and the objective value at 60.3862. Fig. 6 shows how the objective value decreases with the number of iterations.

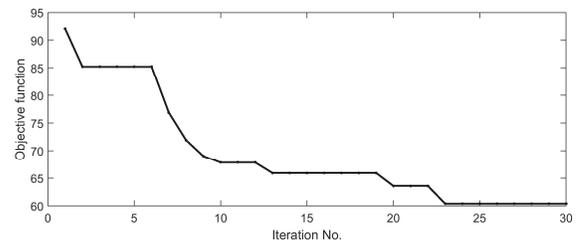

Fig. 6. The iteration curve of the optimal solution for the NE 39-bus system.

### B. Effectiveness under different operating conditions

To investigate the placement results under different operating conditions, the following four scenarios are tested: "**LoadDown**" (active load reduced by 2.5%); "**GenUp**" (active generation increased by 2.5%); "**GenLoadDown**" (both active load and generation reduced by 2.5%); "**GenDownUp**" (a half of the generators increase active powers by 2.5% and the other half of generators decrease by 2.5%). Note that, in practice not all the scenarios need to be run, because in some scenarios the damping ratio of the target mode even without energy storage is already larger than the required damping ratio of 5%.

The considered scenarios and optimal solutions are listed in Table II. As in [30], the **P**lacement **S**imilarity **I**ndex (*PSI*) is calculated for each case, i.e. the ratio of the locations same as those in the base case to all newly selected locations. From the table, the location results are basically consistent. Although $k_{es}$ values vary with the operating condition, in the real life, it is relatively easier to adjust them as controller parameters than to alter those fixed BESS sites when operating condition varies.

TABLE II. OPTIMIZAITON RESULT: DIFFERENT OPERATING CONDITIONS

|  | *locs* | Old $\xi_k$ (%) | New $\xi_k$ (%) | Obj. | PSI |
|---|---|---|---|---|---|
| **LoadDown** | 34 **36 35** | 1.571 | 5.020 | 69.609 | **0.67** |
| **GenUp** | 26 **38 36** | 3.060 | 5.004 | 50.935 | **0.67** |
| **GenLoadDown** | 30 **35 38** | 1.834 | 5.004 | 46.571 | **0.67** |
| **GenDownUp** | **38 35 36** | 2.923 | 5.002 | 60.674 | **1.00** |

### C. Verification of the optimality

To better demonstrate the performance of the optimal solution for the original scenario, the following solutions are tested as shown in Table III: 1) #1 and #2, fixing the ***locs*** vector and randomly varying each $k_{esi}$ within a percentage of −5% to 25%; 2) #3 to # 5, fixing the ***$k_{es}$*** vector and randomly changing each $loc_i$ one-by-one.

The rotor angle curve of generator 1 (g1 in Fig. 5) is considered, which has the highest observability for the target mode on the rotor angles. The comparison results between our optimal solution and solutions with #1 to #5 are shown in Fig. 7, where the highly nonlinear system responses in the first 1 second are ignored, and "w/o" stands for "the case without BESS". From Table III and Fig. 7, the obtained solution can achieve local optimum with satisfactory performance.

TABLE III. SIMULATION VERIFICATION

|  | *locs* | $k_{es}$ | $\Sigma k_{es}$ | $\xi_k$% |
|---|---|---|---|---|
| **Opt.** | **35, 36, 38** | **29.5894, 8.2391, 22.5577** | **60.3862** | **5.005** |
| #1 |  | 28.9957, 8.0738, 22.1051 | 59.1746 | 4.985 |
| #2 |  | 29.9136, 8.3294, 22.8048 | 61.0479 | 5.016 |
| #3 | 13, 36, 38 | **29.5894, 8.2391, 22.5577** | **60.3862** | 4.681 |
| #4 | 35, 27, 38 |  |  | 4.917 |
| #5 | 35, 36, 25 |  |  | 4.727 |

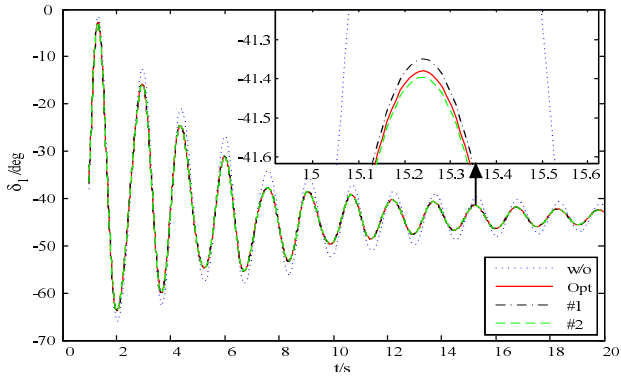

(a) Fixed BESS locations

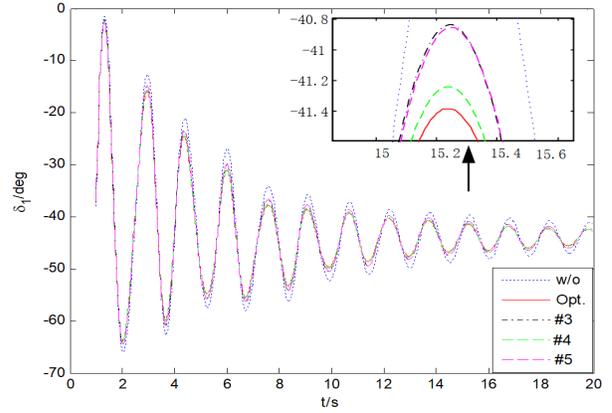

(b) Fixed $k_{es}$ values

Fig. 7. Comparison results on the rotor angle of generator 1.

### D. BESS outputs and SOC changes

Power responses (positive values mean exporting) of BESS units are illustrated in Fig. 8. The change of the SOC for each BESS is calculated by first integrating the response over time and then dividing the result by the capacity of 10 MWh:

- $\Delta SOC_{35}$ = −0.0062 MWh /10 MWh = −0.062 %;
- $\Delta SOC_{36}$ = −0.0017 MWh /10 MWh = −0.017 %;
- $\Delta SOC_{38}$ = −0.0047 MWh /10 MWh = −0.047 %.

Compared with the total energy capacity of each BESS, the final SOC change at the end of oscillation is very small.

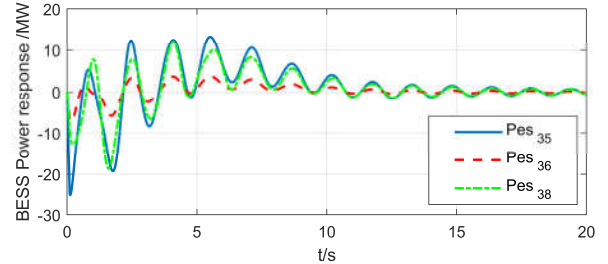

Fig. 8. The BESS power responses in the NE 39-bus system.

## VI. CASE STUDY II: NORDIC TEST SYSTEM

### A. System model and optimization result

The Nordic 20-generator 74-bus test system (operated at 50Hz) models a part of the northern European power grid and consists of four areas as shown in Fig. 9 [31]. The system is heavily loaded with large power transfers from the "**North**" area to the "**Central**" area. Originally, the system frequency is only controlled by the speed governors of the hydro generators in the "**North**" and "**Equiv**" areas. The thermal units of the "**Central**" and "**South**" areas do not participate in this control. "g20" is an equivalent generator with a large participation in primary frequency control.

The original DIgSILENT model has enough damping for each mode. Thus, a modified system was generated by increasing the gains of the AVRs of six generators. Then, an inter-area mode at 0.537Hz with 1.15% damping ratio was selected as the target mode here.

The disturbance considered in simulation is a three-phase temporary fault on Line 4032-4044 at 1 sec and cleared at 1.1 sec. The disturbance is simulated for 50 seconds. This heavy loaded line lies on the critical system oscillation interface.

Parameters of the PSO are: $c_1=2$, $c_2=2$, $\omega=0.9$, $P=30$ (maximum populations) and $I=35$ (maximum iterations), and $k_{es}=[5, 100]$. Still let $N_{es}$ be 3. The best location vector and controller gain vector are: ***locs*** = [g19, g20, 4063] and ***$k_{es}$*** = [81.616, 88.435, 82.564] with damping ratio of the target mode at 5.007% and the objective value at 252.615.

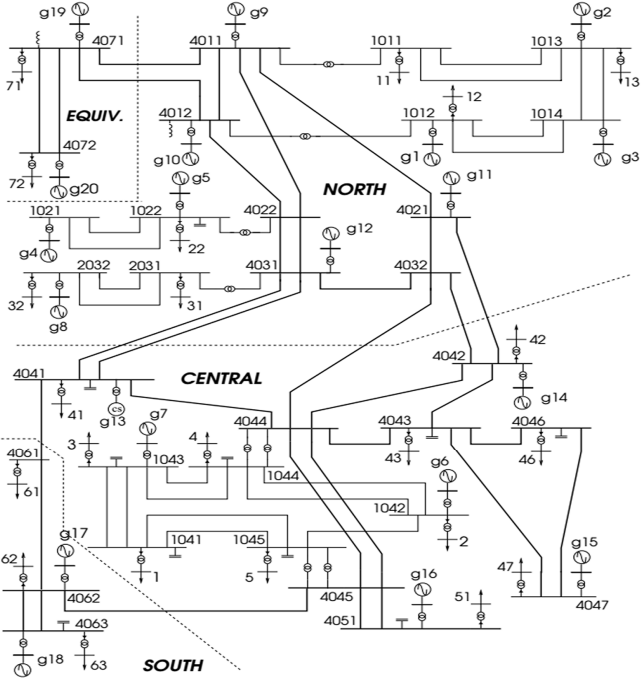

Fig. 9. The Nordic test system diagram.

### B. Effectiveness under different operating conditions

Beside the base condition, three other operating conditions "**LoadDown**", "**GenLoadDown**" and "**GenDownUp**" are tested, similar to those with the NE 39-bus system. The results in Table IV show the robustness of the optimal solution since two of three locations are unchanged.

TABLE IV. OPTIMIZAITON RESULT: DIFFERENT OPERATING CONDITIONS

|  | locs. | Old $\xi_k$% | New $\xi_k$% | Obj. | PSI |
|---|---|---|---|---|---|
| **LoadDown** | g19<br>4063<br>g20 | 2.201 | 5.019 | 189.41 | **1.00** |
| **GenLoadDown** | g20<br>4063<br>g14 | 1.773 | 5.007 | 277.58 | **0.67** |
| **GenDownUp** | 4044<br>4063<br>g20 | 1.170 | 5.031 | 300.00 | **0.67** |

### C. Verification of the optimality

The rotor angle of generator 18 (g18 in Fig. 9) is chosen to illustrate the local optimality of the solution with the base operating condition. It is compared to solutions with slight changes in the ***$k_{es}$*** vector (#1 and #2) and the ***locs*** vector (#3 to #5) as shown in Table V and Fig. 10.

TABLE V. SIMULATION VERIFICATION

| | locs. | $k_{es}$ | $\Sigma k_{es}$ | $\xi_k$% |
|---|---|---|---|---|
| **Opt.** | g19, g20, 4063 | **81.616, 88.435, 82.564** | **252.615** | **5.007** |
| #1 | | 80.849, 87.953, 82.180 | 250.982 | 4.985 |
| #2 | | 83.640, 90.111, 84.209 | 257.960 | 5.086 |
| #3 | g9, g20, 4063 | 81.616, 88.435, 82.564 | 252.615 | 4.558 |
| #4 | g19, 62, 4063 | | | 4.305 |
| #5 | g19, g20, 4 | | | 4.368 |

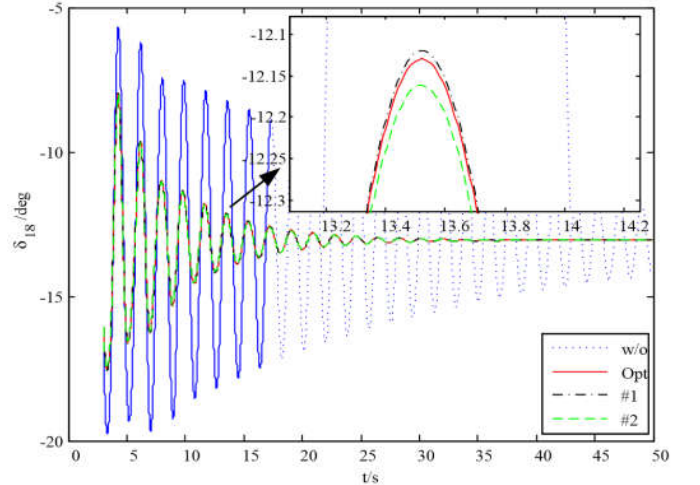

(a) Fixed BESS locations

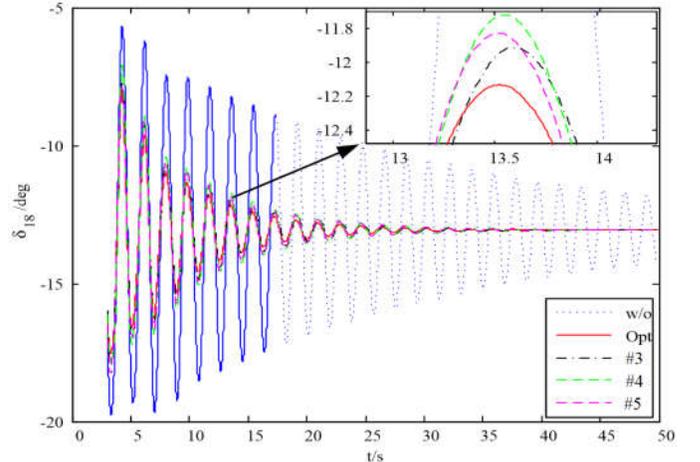

(b) Fixed $k_{es}$ values

Fig. 10. Comparison results on the rotor angle of generator 18

### D. BESS outputs and SOC changes

The power responses of BESS units are shown in Fig. 11. The SOC changes of three BESSs are given below, which are small at the end of oscillation:

- $\Delta SOC_{g19}$ = −0.00021 MWh /10 MWh =−0.0021%
- $\Delta SOC_{g20}$ = −0.00023 MWh /10 MWh =−0.0023%
- $\Delta SOC_{4063}$ = −0.00021 MWh /10 MWh =−0.0021%

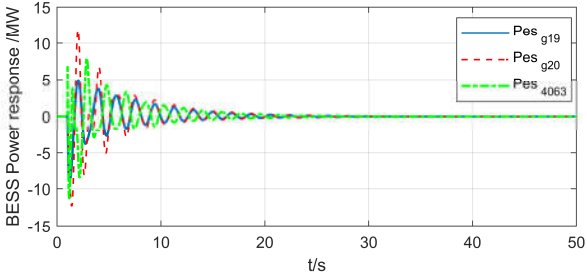

Fig. 11. The BESS power responses in the Nordic test system.

## VII. Applicability Studies

This section further investigates the applicability of the proposed approach and then improves its performance.

### A. Consideration of Seasonal Load Changes

Seasonal load changes can be modeled by a series of different loading levels. In order to accommodate the optimization formulation to multiple loading levels, the following modification is adopted:

$$\min_{z_i, k_{esi}} Obj = \sum_{i=1}^{N} z_i k_{esi} \quad (11)$$

$$\text{s.t.} \quad \sum_{i=1}^{N} z_i = N_{es} \quad z_i \in \{0, 1\} \quad (12)$$

$$\xi_k^l \geq \xi_k^* \quad l \in \{1, 2, ..., L\} \quad (13)$$

$$\Delta \xi_i^l = \xi_i^l - \xi_{i0}^l \geq 0 \quad i = 1, 2, ..., n \quad (14)$$

$$k_{es,\min} \leq k_{esi} \leq k_{es,\max} \quad i = 1, 2, ..., N_{es} \quad (15)$$

where $L$ is the total number of different seasonal loading levels to be considered, $\xi_k^l$ is the damping ratio of the target mode under the $l$-th loading level. Thus, damping ratios of the target mode under all the loading levels will be checked simultaneously. This new formulation is tested on both systems as follows.

For the NE system, assume the normal loading level and a high loading level (+20%). The optimal location and controller gain vectors are found to be ***locs*** = [23, 38, 5] and ***k***$_{es}$ = [24.9072, 38.8458, 5.0603] with damping ratios of the target mode at [5.009%, 5.936%]. The objective function is 68.8133.

For the Nordic system, assume the normal loading level and a high loading level (+10%). The optimal solution gives ***locs*** = [g6, g20, g18] and ***k***$_{es}$ = [95.2295, 97.6079, 92.3342] with the damping ratios of the target mode at [5.303%, 4.992%]. The objective function equals 285.1716.

### B. Determination of the Number of BESSs by Cost Analysis

Generally speaking, the overall investment cost of the placed BESS units depends on both the energy capacities of battery cells and the power capacities of the power electronic converters. The BESS number can be empirically determined by the following two steps:

1) Run the proposed optimization for a reasonable range of $N_{es}$ based on engineering judgement or *a priori* knowledge. For instance, in [14], $N_{es}$ is equal to the number of interested inter-area modes needed for damping improvement. In this paper, $N_{es}$ is examined from 1 to 6 for the two studied systems.

2) Estimate the investment cost based on the optimization results by using the following cost-evaluation model:

$$Cost_{Total} \approx Cost_{Conv} + Cost_{Cell}$$
$$= \sum_{j=1}^{N_{es}} \left( P_{esj} S_{Base} Cost_1 \right) + N_{es} E Cost_2$$
$$= \sum_{i=1}^{N} \left( z_i k_{esi} |\Delta \omega|_{\max} S_{Base} Cost_1 \right) + N_{es} E Cost_2$$
$$= Obj |\Delta \omega|_{\max} S_{Base} Cost_1 + N_{es} E Cost_2 \quad (16)$$

This model prices the power converter cost and battery cell cost separately as two major portions in the overall investment. The other costs, e.g. the installation cost, tax and regulation cost, are assumed to be already contained by those two costs. The meanings of symbols in (16) are: $Cost_{Conv}$ and $Cost_{Cell}$ are respectively the costs for the power converter and battery cells; $Obj$ is the objective value of the optimal solution of problem (4)-(8); $S_{Base}$ is the MVA base equal to 100MVA; $Cost_1$ and $Cost_2$ are respectively the unit costs for 1) the power converter in terms of the power capacity and 2) the battery cells in terms of the energy capacity. They take \$421.43/kW and \$218.52/kWh respectively, inferred from Tesla Powerwall [32]; $E$=10 MWh is the energy capacity of each BESS; $|\Delta \omega|_{\max}$ is the maximum possible frequency deviation (p.u.) to be considered.

A larger $|\Delta \omega|_{\max}$ can lead to a more conservative estimation result for the power capacity of BESS converters while smaller $|\Delta \omega|_{\max}$ may cause insufficient damping support. Since larger frequency deviation will be typically covered by Remedial Action Schemes (RAS) such as Under Frequency Load Shedding and generator tripping schemes [33], the BESS units are expected to help damp oscillations before RAS actions. Moreover, the low frequency oscillation is essentially a small-signal stability problem typically caused by a small disturbance. Therefore, $|\Delta \omega|_{\max}$ = 0.01 p.u. (e.g. 0.6Hz for 60Hz system) is assumed here in the BESS cost evaluation.

The final comparison results are shown in Tables VI and VII: for the NE 39-bus system, two BESS units are enough to provide necessary oscillation damping supports; for the Nordic test system, at least three BESS units are needed.

TABLE VI. NE 39-BUS SYSTEM: COST ANALYSIS FOR DIFFERENT BESS NUMBERS

| $N_{es}$ | $Obj$ | Constraint | $Cost_{Conv}$ ($10^6$ \$) | $Cost_{Cell}$ ($10^6$ \$) | $Cost_{Total}$ ($10^6$ \$) |
|---|---|---|---|---|---|
| 1 | 50.000 | Unsatisfied | 21.072 | 2.1852 | 23.257 |
| 2 | 59.722 | Satisfied | 25.169 | 4.3704 | 29.539 |
| 3 | 60.386 | Satisfied | 25.448 | 6.5556 | 32.004 |
| 4 | 59.766 | Satisfied | 25.187 | 8.7408 | 33.928 |
| 5 | 66.354 | Satisfied | 27.964 | 10.926 | 38.890 |
| 6 | 72.739 | Satisfied | 30.654 | 13.111 | 43.766 |

TABLE VII. NORDIC TEST SYSTEM: COST ANALYSIS FOR DIFFERENT BESS NUMBERS

| $N_{es}$ | $Obj$ | Constraint | $Cost_{Conv}$ ($10^6$ \$) | $Cost_{Cell}$ ($10^6$ \$) | $Cost_{Total}$ ($10^6$ \$) |
|---|---|---|---|---|---|
| 1 | 100.00 | Unsatisfied | 42.143 | 2.1852 | 44.328 |
| 2 | 200.00 | Unsatisfied | 84.286 | 4.3704 | 88.656 |
| 3 | 252.62 | Satisfied | 106.46 | 6.5556 | 113.02 |
| 4 | 301.65 | Satisfied | 127.12 | 8.7408 | 135.86 |
| 5 | 317.51 | Satisfied | 133.81 | 10.926 | 144.74 |
| 6 | 271.03 | Satisfied | 114.22 | 13.111 | 127.33 |

## C. Searching Efficiency Improvement for PSO

The computational complexity of a classic PSO algorithm is in $O(P \times I \times T)$, where $P$ is the population size, $I$ is the maximum iteration number and $T$ is the time cost for objective function evaluation in each time which is roughly equal to the single-run transient simulation time of the power grid.

An alternative analytical approach for BESS placement is to find the locations with the highest modal observabilities regarding the target mode. By small-signal analysis, the normalized modal observabilities on generator speeds are listed in Table VIII for the NE 39-bus system.

TABLE VIII.  NE 39-BUS SYSTEM: TARGET MODE OBSERVABILITIES ON GENERATOR SPEEDS

| Gen No. | 9 | 5 | 6 | 7 | 4 |
|---|---|---|---|---|---|
| Bus No. | 38 | 34 | 35 | 36 | 33 |
| Observability | 0.4109 | 0.4008 | 0.3681 | 0.3656 | 0.3508 |
| Gen No. | 3 | 2 | 8 | 10 | 1 |
| Bus No. | 32 | 31 | 37 | 30 | 39 |
| Observability | 0.2710 | 0.2514 | 0.2292 | 0.2250 | 0.1962 |

However, sequentially picking the top-$N_{es}$ locations based on the above observabilities might not guarantee the optimality of the locations. The reason is that the proposed simulation-based optimization uses detailed simulation models. For example, suppose that all $k_{es,i}$ have the same value, e.g. 20 for the NE system; then using the locations [38, 34, 35] with top-3 observabilities for BESS placement, the damping ratio of the target mode is 5.169%, lower than damping ratio 5.221% with the previously optimized locations [35, 36, 38] having $k_{es,i}$=20. Thus, the proposed simulation-based optimization approach can give more credible optimal solutions than the aforementioned analytical approach.

From the optimization results in Sections V and VI, although the optimal locations consist of both generator buses and non-generator buses, the locations are often on or close to generator buses. Thus, to enhance the searching efficiency of the proposed Mixed-PSO algorithm, the candidate-bus set is reduced to the set of all generator buses and the top-$m$ closest buses to each generator bus, so as to reduce the population size $P$ of the PSO method. In the following, $m$=1 is considered.

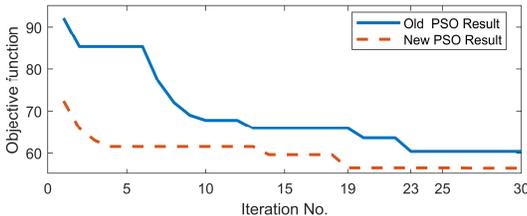

Fig. 12. Comparison of the original (solid) and improved (broken) PSOs.

For the NE 39-bus system, still let $N_{es}$ = 3 and run the improved PSO procedure only for reduced candidate buses. The comparison plots for the iteration curves are shown in Fig. 12. It is obvious that the improved PSO algorithm converges earlier at the 19th iteration than before, i.e. at the 23rd iteration. For the Nordic test system, a similar test is done.

The test results on both systems are summarized in Table IX, where the "Old result" stands for the baseline result in Sections V and VI and the "New result" is obtained by the improved PSO method. It can be observed that using generators and nearby buses to initialize the Mixed-PSO population provides a good initial solution, especially for the discrete part (i.e. locations) of each particle. This measure helps to speed up the convergence of the searching process. Regarding the solution optimality, for NE 39-bus system, this new scheme achieves a smaller objective function value. For a larger system like the Nordic system, it yields a slightly bigger objective function value though still close to the old result.

TABLE IX. COMPARISON OF OLD PSO AND IMPROVED PSO ALGORITHMS

| NE 39-bus | Locs. | $\xi_k$% | Obj. |
|---|---|---|---|
| Old result | 35, 36, 38 | 5.005 | 60.3862 |
| New result | 33, 35, 38 | 5.000 | 56.4446 |
| **Nordic test** | Locs. | $\xi_k$% | Obj. |
| Old result | g19, g20, 4063 | 5.007 | 252.615 |
| New result | g16, g20, 4063 | 5.009 | 265.923 |

## D. Controller Performance Comparison

In this sub-section, a preliminary comparison against the following controller inspired by [23] is conducted under the assumption of pre-determined three locations.

$$K(s) = k_{es} + \frac{1}{T_i s} \qquad (17)$$

In Section V, the optimal location vector and controller gain vector are **locs** = [35, 36, 38] and $\boldsymbol{k_{es}}$ = [29.5894, 8.2391, 22.5577] with damping ratio of the target mode at 5.005%. Then, set $T_i$ of the above PI controller at 0.01, 0.1 and 1 (s) for a comparison with the proposed controller defined by (2) and (3) in this paper. The resultant damping ratios are respectively 3.826%, 4.055% and 5.151%. With an integrator added, the performance of the PI controller in terms of the damping ratio of the target mode becomes worse when $T_i$ is small. Although a larger $T_i$ may bring certain improvement, the price is the much larger BESS power overshoot during the transient period as illustrated in Fig. 13. As a result, that will cause higher power capacity ratings for power converters and hence higher investment costs, which is undesired.

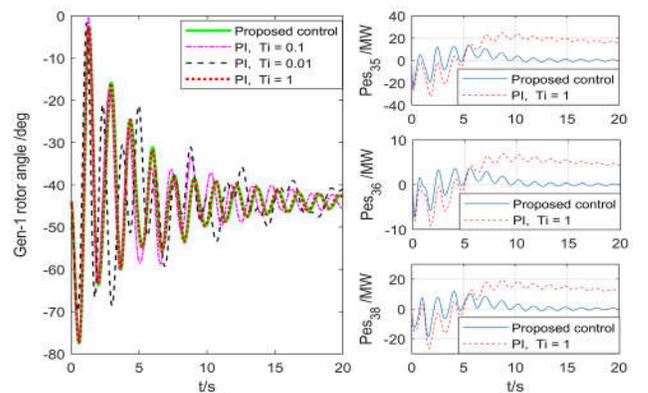

Fig. 13. Performance comparison of different controllers.

## VIII. Conclusion

This paper has proposed a simulation-based optimization approach for the optimal placement and control parameter settings of multiple BESS units to improve oscillation damping in a power transmission system. The approach employs a Mixed-Integer PSO method to solve the optimization problem and is tested on two power systems. The optimality of the given optimal solution and the impacts of operating conditions on the solution are studied. The proposed optimization scheme can accommodate seasonal load changes and can be applied for cost analysis regarding BESS units. The controller assumed in the BESS model is compared with another typical type of controllers in the existing literature to validate its superiority.

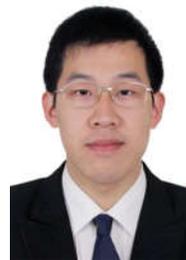

**Yongli Zhu** (S'12) received his B.S. degree from Huazhong University of Science & Technology in 2009 and M.S. degree from State Grid Electric Power Research Institute in 2012. He is now pursuing the Ph.D. degree at the University of Tennessee, Knoxville. His research interests include energy storage and artificial intelligence for power system stability and control.

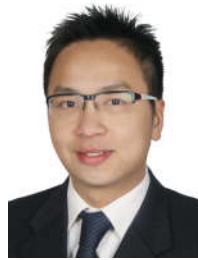

**Chengxi Liu** (S'10-M'13) received his B. Eng. and M. Sc. degrees in Huazhong University of Science and Technology, China, in 2005 and 2007 respectively. He received the Ph.D. degree at the Department of Energy Technology, Aalborg University, Denmark in 2013. He worked in Energinet.dk, the Danish TSO, until 2016. He was a Research Associate at the Department of EECS, the University of Tennessee, USA until 2018. Currently, he is an Associate Professor at the Department of Energy Technology, Aalborg University, Denmark.

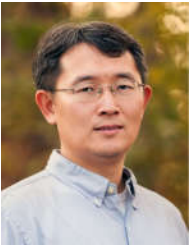
**Kai Sun** (M'06–SM'13) received the B.S. degree in automation in 1999 and the Ph.D. degree in control science and engineering in 2004 both from Tsinghua University, Beijing, China. He is currently an associate professor at the Department of EECS, University of Tennessee, Knoxville, TN, USA. He was a project manager in grid operations and planning at the EPRI, Palo Alto, CA from 2007 to 2012. Dr. Sun serves in the editorial boards of IEEE Transactions on Smart Grid, IEEE Access and IET Generation, Transmission and Distribution. His research interests include stability, dynamics and control of power grids and other complex systems.

**Di Shi** (M'12, SM'17) received the Ph.D. degree in electrical engineering from Arizona State University, Tempe, AZ, USA, in 2012. He currently leads the PMU & System Analytics Group at GEIRI North America, San Jose, CA, USA. Prior to that, he was a researcher at NEC Laboratories America, Cupertino, CA, and Electric Power Research Institute (EPRI), Palo Alto, CA. He served as Senior/Principal Consultant for eMIT and RM Energy Marketing between 2012-2016. He has published over 70 journal and conference papers and holds 14 US patents/patent applications. He received the IEEE PES General Meeting Best Paper Award in 2017. One Energy Management and Control (EMC) technology he developed has been commercialized in 2014 into product that helps customers achieve significant energy savings. He is an Editor of IEEE Transactions on Smart Grid.

**Zhiwei Wang** received the B.S. and M.S. degrees in electrical engineering from Southeast University, Nanjing, China, in 1988 and 1991, respectively. He is President of GEIRI North America, San Jose, CA, USA. His research interests include power system operation and control, relay protection, power system planning, and WAMS.